\documentstyle[prb,psfig,epsfig,aps]{revtex}

\input{epsf}

\begin{document}
\draft
\title {Symmetry broken motion of a periodically driven Brownian particle: 
nonadiabatic regime.
}

\author{M. V. Fistul}
\address{Max-Planck Institut f\"ur Physik Komplexer Systeme,
D-01187, Dresden, Germany}
\date{\today}

\wideabs{ %REVTeX 3.1 feature

\maketitle

\begin{abstract} We report a theoretical study of an overdamped
Brownian particle dynamics in the presence of both  
a spatially modulated one-dimensional periodic 
potential $U(x)$ and  
a periodic alternating force (AF). As the periodic 
potential $U(x)$ has a low 
symmetry (a {\it ratchet potential}) the Brownian particle displays 
a broken symmetry motion with a nonzero time average velocity.
By making use of the Green function method and a mapping to the 
theory of Brillouin bands the probability distribution $P(x,t)$ of 
the particle coordinate $x$ is derived and  
the nonlinear dependence of the macroscopic velocity on the 
frequency $\omega$ and the amplitude $\eta$ 
of AF is found. In particular,  
our theory allows to go beyond the adiabatic limit ($\omega~=~0$) and to 
explain the peculiar reversal of the velocity sign found 
previously in the numerical analysis.
\end{abstract}

}

%end of WideAbs
%\newpage

The model of a particle moving in a periodic one-dimensional potential $U(x)$ 
has been used to explain diverse
fascinating phenomena in various fields of physics, chemistry and biology
\cite{Strogatz,Stratan,Risken}.
Well known examples of such phenomena are Brillouin bands and 
Bloch oscillations in the solid state theory \cite{Kittel}, 
dynamics of Josephson junctions \cite{Likh}, energy transport in various 
biological systems \cite{Prost}, etc.

The peculiar property of a particle motion in 
the periodic potential is the presence of two
different states, a static state and a whirling (dynamic) state. 
In the dynamic state, the particle has a non zero value of the
time averaged velocity $v=<\dot x(t)>$. 
As the system is subject to an external noise, the 
"{\it Brownian}" particle motion  
becomes more complex (it displays both random damped oscillations 
and random jumps between the potential wells \cite{vanKampen}), 
the difference between libration and rotation states
disappears. 

The theory of Brownian particle motion in a specific periodic 
potential of $ \cos $-like shape and in the presence of a constant driving force 
(DF) has been developed in Refs. 
\cite{Stratan,IZ,AH}. As a particular result it was found that in the 
absence of DF the mean particle 
velocity is zero. Moreover,  
it is well known \cite{Fein,Prost} that 
equilibrium noise only can not lead to
the fluctuation induced transport, and correspondingly, 
$v~=~0 $ for the particle motion in 
an arbitrary periodic potential $U(x)$. In this case the diffusive  
motion of particle occurs. The probabilities of 
fluctuation induced particle escape from the potential well to the left 
and to the right are 
identical. These probabilities are determined by the amplitude of
potential $U(x)$ and do not depend on the symmetry of potential. 

The situation changes drastically when a Brownian particle is subject to 
both {\it a periodic spatially modulated potential (PP) and 
a periodic alternating force 
(AF)}. It was shown in Ref. \onlinecite{Hangi} by numerical analysis of 
the corresponding 
Fokker-Planck equation that in this case the reflection symmetry of an 
overdamped 
particle motion ($x$ to $-x$) can be 
broken, and directed transport occurs. 
The specific condition of such transport is
the low symmetry of the periodic potential $U(x)$ (a {\it ratchet potential}).
Moreover, by making use of the symmetry arguments and the 
numerical analysis it was shown  that directed transport occurs 
even in the presence of symmetric PP but as the periodic AF has a low symmetry in time 
\cite{Fluch}.

An analytical description of a Brownian particle motion in an anisotropic 
periodic potential has been carried out in an adiabatic regime as   
the frequency of AF $\omega$ is rather small \cite{Hangi,Landa}. 
In this limit it was found for 
arbitrary values 
of the noise strength and the amplitude of AF $\eta$, 
the particle moves in the direction 
of the slower rate of potential change \cite{Landa}. 
However, in Ref. \onlinecite{Hangi} the peculiar {\it reversal of the sign} 
of $v$ has been found
as the AF frequency $\omega$ becomes relatively large.
In this nonadiabatic limit and as the amplitude of AF $\eta$ increases, 
the dependence $v(\eta)$ displays oscillations. 
Moreover, these effects do not appear in the overdamped deterministic 
regime where the noise strength is zero \cite{comment2}. Notice here,
 that the directed transport of Brownian particle is more complex 
than the one in the deterministic case.
In latter case, the non zero value of $v$ appears as a simple 
consequence of two different values of a depinning force, and the 
directed transport is absent as the amplitude of AF is small.    
Thus, in order to explain various 
peculiarities of the overdamped Brownian particle motion 
we develop a theoretical analysis that goes beyond an adiabatic approximation.  

In this Letter we present a consistent analytical approach to the dynamics 
of an overdamped Brownian particle in the 
presence of both PP $U(x)$ with the period $a$ and harmonic AF,   
$\eta \cos (\omega t)$ \cite{comm1}. By making use of a particular 
diagrammatic technique that is valid in the limit of a large time, 
the dependence of the mean velocity on the 
amplitude $\eta$ and frequency $\omega$ of AF will be calculated.
We find that in the presence of AF the average velocity is determined by 
various {\it relaxation processes inside the potential well}, 
and the directed transport 
of Brownian particle occurs as the potential $U(x)$ has a 
low symmetry. We also obtain 
the particular range of $\omega$ where the interplay 
between noise and AF leads to the reversal of the sign of $v$. 

The dynamics of an overdamped particle is described by the Langeven equation
\cite{vanKampen} 
:
\begin{equation} \label{LangEq}
\alpha \dot x(t) + U' (x) ~=~ \eta \cos (\omega t)+\xi (t)~~,
\end{equation}
where the white noise function $\xi (t)$ has a zero mean and the 
correlation function $<\xi (t) \xi ( t')>~=~2\alpha T \delta(t- t')$. 
Here, $\alpha$ is the damping coefficient and $T$ is the effective temperature 
describing the strength of the fluctuations.
Next, we introduce the time-dependent probability density $P(x,t)$ that 
satisfies the Fokker-Plank equation \cite{Stratan,Risken}:
\begin{equation} \label{FPeq}
\alpha \frac{\partial P(x,t)}{\partial t} ~=~ T\frac{\partial^2 P(x,t)}{\partial x^2}+
\frac{\partial}{\partial x}\{[ U' (x)-\eta \cos (\omega t)]P(x,t)\}~~~
\end{equation}
with the initial condition 
\begin{equation} \label{Incond}
P(x,0) ~=~\delta(x-x_0) ~~.
\end{equation}
Using the Laplace transformation $P_\lambda (x) ~=~\int_0^\infty dt 
P (x,t)e^{-\lambda t}$ 
and the standard substitution $P_\lambda(x,x_0)~=~e^{-\frac{U(x)-U(x_0)}{2T}} 
G_\lambda(x,x_0)$ 
we obtain the integral 
equation for the Green function $G_\lambda(x,x_0)$ 
$$
G_\lambda(x,x_0) ~=~ G_{\lambda}^{0}(x,x_0)+\int \frac{d\lambda_1}{2\pi i} 
\eta \frac{2\lambda_1}{\omega^2+\lambda_1^2} 
$$
\begin{equation} \label{Intequation}
\frac{1}{\alpha}\int dy G_{\lambda}^{0}(x,y) e^{\frac{U(y)}{2T}}\frac{d}{dy} 
[e^{-\frac{U(y)}{2T}} 
G_{\lambda-\lambda_1}(y,x_0)]~~,
\end{equation}
where $G_{\lambda}^{0}(x,x_0)$ is the Green function of the equation:
$$
T\frac{d^2}{dx^2}G_\lambda^{0}(x,x_0)+G_\lambda^0(x,x_0)(-Te^{\frac{U(x)}{2T}}\frac{d^2}{dx^2}e^{-\frac{U(x)}{2T}} -
\alpha \lambda)~=~
$$
\begin{equation} \label{NofieldGF}
 =-\alpha \delta(x-x_0)~~~~~~~~~~~~~~~~~~~~~~~~~~~.
\end{equation}
The mean value of particle velocity $v$ is determined by the probability 
$P(x,t)$
in the limit of large time:
\begin{equation} \label{Veldef}
v~=~\lim_{t \to \infty} \int dx P(x,t) \frac{(x-x_0)}{t}~~.
\end{equation}
In the absence of AF ($\eta~=~0$) it 
immediately follows from Eq. (\ref{NofieldGF}) and the property 
\cite{Kurant} 
$
G_\lambda^0(x,x_0)~=~G_\lambda^0(x_0,x)~~
$ that the mean value of the velocity is zero 
for an arbitrary periodic potential $U(x)$. However, in the presence of AF the 
probability density $P(x,t)$ is determined by the product of the 
Green functions with different arguments $\lambda$. It leads to the breaking 
of the symmetry and, therefore, to the directed transport.

To obtain the fluctuation induced transport in the presence of AF 
we need to know  the long time 
behaviour of the probability 
distribution $P(x,t)$ (see, the Eq. (\ref{Veldef})). This limit corresponds 
to small values of $\lambda$ in the integral equation
(\ref{Intequation}). Physically it may be interpreted as the particle return to 
the stable state after the multiple interactions with AF. 
Moreover, we are interested in a 
non oscillating part of $P(x,t)$ only. It allows greatly to simplify the 
Eq. (\ref{Intequation}) just keeping most singular contributions 
with small
"lambda" arguments of the 
Green functions $G^{0}_\lambda (x,y)$. These terms 
can be presented in the diagrammatic 
form (see, Fig. 1). By summing these terms and 
calculating the intermediate integrals over 
$\lambda_n$  we obtain the expression 
for $G_\lambda(x,x_0)$ in the form:
\begin{equation} \label{Gexp}
G_\lambda (x,x_0)~=~\int dp e^{ip(x-x_0)} G_\lambda^{0}(p;x,x_0) 
\frac{Z_\lambda(p)}{1-Z_\lambda(p)}~~.
\end{equation}
Here, we introduce the {\it periodic} Green function $G_\lambda^0(p;x,x_0)$ 
written in the mixed momentum -coordinate representation, and
$Z_\lambda(p)$ is determined by expression:
$$
Z_\lambda (p)~=~\frac{\eta^2}{\alpha^2}
\int dy_1 dy_2 P^{0}_{\lambda}(p;y_2,y_1)
$$
\begin{equation} \label{Zexp}
\frac{d}{dy_1}\frac{d}{dy_2}
Re~P_{i\omega}(y_1,y_2)~~.
\end{equation}

\begin{figure}
  \centering
  \psfig{figure=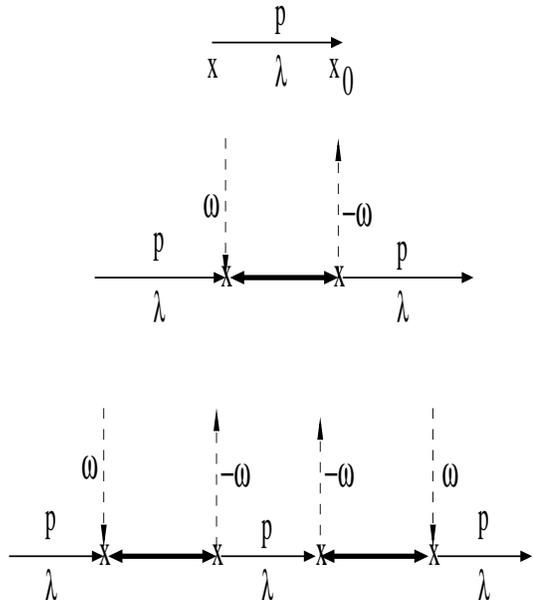 ,height=8cm ,width=7cm}
  \caption{The most important diagrams and 
   the diagrammatic presentation of Eq. (\ref{Intequation}). Thin and thick 
   solid lines correspond accordingly 
   to $G_\lambda^0(p;x,y)$ and $P_\lambda(x,y)$; dashed lines are due to the 
   presence of AF. The cross presents an operator 
   $\frac{\eta}{\alpha} \frac{d}{dx}$. In the limit of large time 
   the parameters $p$ and 
   $\lambda$ are small 
   }
\end{figure}

To calculate the function $Z_\lambda (p)$ and correspondingly, to obtain 
the conditions when the directed transport occurs, we use the 
general properties 
of the Eq. (\ref{NofieldGF}).  This equation can be mapped to a well 
known problem of an electron motion in a periodic potential 
\cite{Risken,Kittel}. Thus, the relaxation times spectrum 
$\frac{\alpha}{\tau_n(p)}$ of 
the Eq. (\ref{NofieldGF}) contains an infinite 
number of bands and is determined by the wave vector $p$.
The lowest band starts from the zero value,  
and in the region of small wave vectors $p$ has a form:
\begin{equation} \label{LowestSpectr}
\frac{1}{\tau_n(p)}~=~ D p^2,~~~p~\ll~\frac{2\pi}{a}~~,
\end{equation} 
where the {\it diffusion coefficient} $D$ depends 
on the strength of the potential and 
temperature. In the limit of small fluctuations as 
the amplitude of $U(x)$ is large and the temperature $T$ is small, 
we obtain that the value of 
$D~\propto~e^{-\frac{U_{max}-U_{min}}{T}}~=~e^{-\frac{\Delta U}{T}} $ 
is also small. 
Note here, that the diffusion coefficient $D$ 
does not depend  
on the symmetry of the potential $U$ and is determined by the  
the narrow regions close to the top and bottom of potential. 
In the absence of AF the  particle displays a diffusive motion and it is 
completely determined by $D$. 
The lowest band 
can be also mapped to a well known tight-binding model \cite{Kittel}.
The corresponding eigen function $\psi_p(x)$ has the form of a Bloch wave 
\cite{Kittel}, namely
$\psi_p(x)~=~e^{ipx}r_p(x)$, where the periodic function 
\begin{equation} \label{LowestEugFunct}
r_p(x)~\simeq~ 
\exp(-\frac{U(x)}{2T})+pu_0(x)
\end{equation}
in the limit of small $p$. Here, 
the first correction $u_0(x)$ in the region of 
small $p$ has been found in Refs. \cite{Risken,Kulik}. 
Upper bands correspond to much smaller relaxation times $\tau_n$ but 
exactly these bands determine the directed transport.
%Thus,
%$\epsilon_n(0) ~\simeq~n^2 T(\frac{2\pi}{a})^2$.

In the limit of large time the main contribution to the probability 
distribution $P(x,t)$ results from a small momentum $p$. The 
presence of an {\it odd} component of 
$Z_{\lambda}(p)~\propto~p$ leads to a nonzero value of $v$, 
and thus, by using (\ref{LowestSpectr}), 
(\ref{LowestEugFunct}) and substituting (\ref{Zexp}) and (\ref{Gexp}) in 
(\ref{Veldef}) we arrive at the expression for $v$
$$ 
v~=~\frac{\eta^2}{\alpha^2} 
\frac{2a}{3A}\int_0^a dy_1\int_0^a dy_2 
e^{\frac{-3U(y_1)}{2T}}e^{\frac{3U(y_2)}{2T}}
$$
\begin{equation} \label{FinVelExp}
\frac{d}{dy_1}
Re~\tilde G_{i\omega}(y_1,y_2)~~,
\end{equation}
where $ A~=~\int_0^a dy_1 \int_0^a  dy_2 
e^{\frac{-U(y_1)}{T}}e^{\frac{U(y_2)}{T}}$. Here, the Green function 
$\tilde G_\lambda (x,x_0)$ 
is a solution of the Eq. (\ref{NofieldGF}) with  
periodic boundary conditions. 
%Acknowledgments

The formula (\ref{FinVelExp}) is convenient for analysis of both 
general properties of 
Brownian particle motion and particular limits. 
Thus, a general feature of an overdamped Brownian particle transport is that 
the average velocity $v$ is determined by the properties of 
potential $U(x)$ on the {\it whole range} of its variation. Moreover, 
the directed transport of Brownian particle 
is an interplay of two effects:
the fluctuation induced particle escape from the potential well (the 
coefficient $1/A$ in the Eq. (\ref{FinVelExp})) and the AF induced 
relaxation processes 
inside the potential well (the Green function $\tilde G_{i\omega}(y_1,y_2)$ 
in the Eq. (\ref{FinVelExp})). 
However, if the potential $U(x)$ displays a reflection symmetry, namely 
$U(x)$ is an even function of $x$, the average value of $v$ vanishes due to 
a fact that $\frac{d}{dy_1}P_{i\omega}(y_1,y_2)$ is an odd function of $y_1$ or 
$y_2$, and one of the integrals 
in the Eq. (\ref{FinVelExp}) is zero. 

If the potential $U(x)$ has no reflection symmetry, the average velocity 
is not zero, and it increases as $\eta^2$ in the limit of a small amplitude 
of AF $\eta$. In this limit by making use of the eigen functions $\psi_n(x)$ 
of the Eq. (\ref{NofieldGF}) we get:
$$ 
<v>~=~\frac{\eta^2}{\alpha^2} 
\frac{2a}{3A}\sum_n \frac{\tau_n}{1+(\omega \tau_n)^2}
$$
\begin{equation} \label{PsiVel}
\int_0^a dy_1\int_0^a dy_2 
e^{\frac{-3U(y_1)}{2T}}e^{\frac{3U(y_2)}{2T}}\psi_n(y_2)\psi_n^{'}(y_1)~~.
\end{equation}
Thus,
the average velocity $v$ is small in both limits of small and large 
effective temperature $T$:
\begin{equation} \label{VelLimit}
v(T) ~\propto~ \left\{
\begin{array}{rl}
e^{-\frac{\Delta U}{T}}, & T~\ll~U_{max}-U_{min} \nonumber \\  
T^{-3}f(\omega \tau_1)~~, & T~\gg~U_{max}-U_{min}~~.
\end{array}
\right.
\end{equation}
The particular function $f$ depends on the ratio between the frequency 
$\omega$ and $\tau_1^{-1}$ where $\tau_1$ is the maximum relaxation time 
of the particle in the potential $U(x)$. 
The Eq. (\ref{PsiVel}) shows that 
the adiabatic regime is valid only in the limit of $\omega~\ll~1/\tau_1$, 
The relaxation time $\tau_1$ depends crucially on the 
effective temperature $T$. Thus, as 
the fluctuations are weak ($T~\ll~\Delta U$),  
$\tau_1~=~\frac{\alpha}{\omega_0^2}$, where $\omega_0$ is the frequency of
 small oscillations at the bottom of potential well \cite{Klim}. 
Moreover, $\tau_1$ can be even larger in the limit of 
small fluctuations 
as the potential $U(x)$ has 
a form of a double potential well \cite{Klim,Gamm}. 
In the opposite regime of large fluctuations ($T~\gg~\Delta U$) the relaxation 
time depends on $T$ and decreases as 
\begin{equation} \label{tau:LT}
\tau_1~=~\frac{\alpha a^2}{4\pi^2 T}~~.
\end{equation} 

In the limit of a small frequency $\omega$
the main contribution to the directed transport results from 
the first "relaxation time" band. It leads to a particular sign of $v$ that 
does not change with the temperature. 
However, as the frequency increases ($\omega~\gg~\tau_1^{-1}$) the upper 
"relaxation time" bands dominate and the sign reversal can occur. 
As T increases the relaxation time $\tau_1$ decreases and the adiabatic limit is
recovered.

This general scenario can be verified in the limit of large temperature 
$T~\gg~
\Delta U$ for a particular 
model of low symmetry potential $U(x)~=~U_0(\cos (\frac{2\pi x}{a})+
\gamma \sin (\frac{4\pi x}{a}))$, where the parameter $\gamma$ is of order one.
As the temperature is large the eigen functions $\psi_n(x)$ of the Eq. 
(\ref{NofieldGF}) are plain waves, the main contribution to the $v$ results 
from the first and second relaxation time bands,
and we obtain:
\begin{equation} \label{Vel:LT}
v(\omega)~=~\gamma\frac{\eta^2}{\alpha^2 a} \frac{U_0^3}{T^3}\tau_1[\frac{1}{1+(\omega \tau_1)^2}-
\frac{4}{16+(\omega \tau_1)^2}]~~, 
\end{equation}
where $\tau_1$ is determined by the Eq. (\ref{tau:LT}).
The dependence of $v(\omega)$ that displays a sign reversal, is presented in 
Fig. 2. 

In the discussion presented to this point we assume that the amplitude of 
AF $\eta$ is small. As $\eta$ increases the peculiar oscillations appear in 
the dependence of $v(\eta)$. The period of these oscillations $\Delta \eta$
can be estimated 
in the same limit of large temperature $T$. By making use of the 
Eq. (\ref{FPeq}) and a smallness of the potential $U$ we find:
\begin{equation} \label{Oscil}
\Delta \eta (T) ~\propto~ \left\{
\begin{array}{rl}
\alpha a \omega~~, & \omega~\gg~\tau_1^{-1} \nonumber \\  
\alpha a \tau_1^{-1}~~, & \omega~\ll~\tau_1^{-1}.
\end{array}
\right.
\end{equation}
\begin{figure}
  \centering
  \psfig{figure=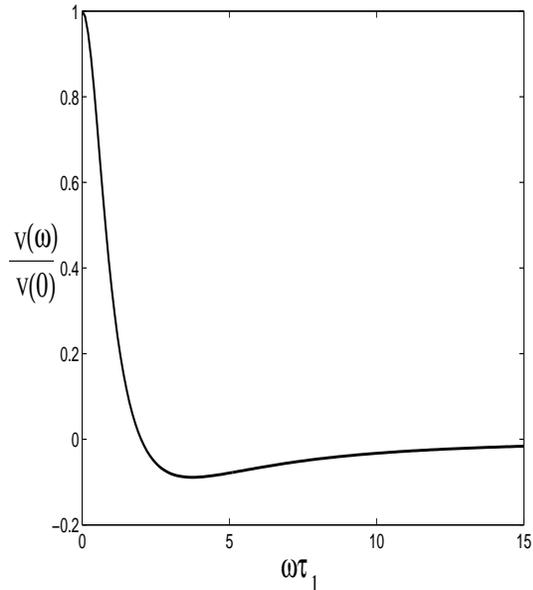 ,height=8cm ,width=7cm}
  \caption{
   The frequency dependence of the mean value of velocity $v$: the limit of 
   large effective temperature.
    }
\end{figure}
In conclusion, we have presented an analysis of an overdamped Brownian 
particle motion in the presence of periodic potential and AF. We have shown 
that the directed transport in such a system is determined by two effects: 
the particle escape from a potential well and the relaxation processes inside 
the potential well. It is at variance with both 
the standard diffusive motion in the absence of AF and 
the nonlinear dynamic response of the system to a weak AF 
\cite{vanKampen,Klim,Gamm}. We argue that the $\omega$ dependent 
reversal of the sign of $v$ 
found previously by numerical analysis is due to different strength and 
symmetry of relaxation processes.  
Thus, the observation of directed transport and especially its dependence on 
the frequency $\omega$ and amplitude $\eta$ of AF, allows to investigate 
the various relaxation processes of an overdamped Brownian particle.
Moreover, the used method of taking into account the most singular diagrams 
where the momentum $p$ and parameter $\lambda$ are small, can be also useful 
in order to analyze the fluctuation induced transport in the presence 
of more complex (periodic or random) AF \cite{Fluch,comm1}. 
   
I thank
S. Flach, O. Yevtushenko, Y. Zolotaryuk and A. V. Ustinov  
for useful discussions.  
%%%%%%%%%%%%%%%%%%%%NEW%%%%%%%%%%%%%%%%%%%%%%%%%%%%%%%%%%%%%%%%%%%%%%%
%%%%%%%%%%%%%%%%%%%%%%%%END NEW%%%%%%%%%%%%%%%%%%%%%%%%%%%%%%%%%%%


\begin{references}
\bibitem{comm1} The fluctuation induced transport in the presence of periodic 
but nonharmonic AF will be reported elsewhere.
\end{references}
\end{document}